\documentclass{article}
\usepackage[a4paper, margin=1in]{geometry}
\usepackage{amsmath}
\usepackage{xcolor}
\usepackage{graphicx}
\usepackage{hyperref}
\usepackage{multirow}
\usepackage{longtable}
\usepackage{pifont}

\newcommand{\RAt}[1]{\textnormal{R@#1}}
\usepackage[textsize=tiny]{todonotes}

\definecolor{red}{RGB}{192,0,0}
\renewcommand{\epsilon}{\varepsilon}

\title{ReSearch: A Multi-Stage Machine Learning Framework for Earth Science Data Discovery}

\author{Youran Sun\thanks{Equal contribution.} \\
Department of Mathematics\\
University of Maryland, College Park, MD, USA\\
\texttt{sun1245@umd.edu}\\ \\
Yixin Wen\footnotemark[1] \\
Department of Geography\\University of Florida, Gainesville, FL, USA\\
\texttt{yixin.wen@ufl.edu} \\ \\
Haizhao Yang\thanks{Corresponding author.} \\
Department of Mathematics\\
Department of Computer Science\\
University of Maryland, College Park, MD, USA \\
\texttt{hzyang@umd.edu} \\
}

\begin{document}

\maketitle

\begin{abstract}
    The rapid expansion of Earth Science data from satellite observations, reanalysis products, and numerical simulations has created a critical bottleneck in scientific discovery, namely identifying relevant datasets for a given research objective.
    Existing discovery systems are primarily retrieval-centric and struggle to bridge the gap between high-level scientific intent and heterogeneous metadata at scale.
    We introduce \textbf{ReSearch}, a multi-stage, reasoning-enhanced search framework that formulates Earth Science data discovery as an iterative process of intent interpretation, high-recall retrieval, and context-aware ranking.
    ReSearch integrates lexical search, semantic embeddings, abbreviation expansion, and large language model reranking within a unified architecture that explicitly separates recall and precision objectives.
    To enable realistic evaluation, we construct a literature-grounded benchmark by aligning natural language intent with datasets cited in peer-reviewed Earth Science studies.
    Experiments demonstrate that ReSearch consistently improves recall and ranking performance over baseline methods, particularly for task-based queries expressing abstract scientific goals.
    These results demonstrate the importance of intent-aware, multi-stage search as a foundational capability for reproducible and scalable Earth Science research.
\end{abstract}

\section{Introduction}

Earth Science research seeks to understand complex, interconnected processes governing the atmosphere, oceans, land surface, cryosphere, and their interactions with human systems.
This endeavor increasingly relies on large-scale observational datasets, reanalysis products, and numerical simulations produced by a diverse ecosystem of satellites, in situ instruments, and computational models.
While the volume and diversity of Earth Science data have grown rapidly, the ability to efficiently identify, access, and integrate relevant datasets has not kept pace.
As a result, data discovery has emerged as a critical bottleneck that constrains scientific productivity, reproducibility, and participation.

Efficient discovery of relevant Earth Science data is impeded by three interrelated challenges.
First, datasets are distributed across repositories with inconsistent metadata schemas, causing identical variables to be labeled differently across sources.
Second, a persistent semantic gap exists between high-level research objectives and the technical descriptors used in metadata, requiring substantial domain expertise to bridge.
Third, as archives scale to petabytes, balancing recall and precision becomes increasingly difficult, demanding reasoning capabilities beyond simple keyword matching.
These challenges are examined in detail in Section~\ref{sec:problem}.

Recent advances in machine learning offer new opportunities to address these limitations.
Semantic embedding models and large language models (LLMs) have demonstrated strong capabilities in representing meaning, handling linguistic variability, and interpreting natural language queries.
These tools provide a promising foundation for bridging the gap between scientific intent and technical metadata.
However, purely neural approaches often lack transparency, stability, and explicit control over recall and relevance.
These limitations indicate the need for discovery frameworks that combine machine learning with principled search strategies, explicitly encoding the stages of reasoning that underlie scientific inquiry.

In this work, we introduce \textbf{ReSearch}, a multi-stage framework for Earth Science data discovery.
ReSearch decomposes the search process into query understanding, high-recall retrieval, and context-aware reranking, combining lexical matching, semantic embeddings, and large language model reasoning to address metadata heterogeneity and the semantic gap.
The main contributions are as follows.

\begin{itemize}
    \item We propose a multi-stage search architecture that explicitly separates intent interpretation, candidate retrieval, and relevance ranking, enabling scalable discovery while preserving scientific relevance.

    \item We develop a hybrid retrieval strategy integrating BM25, semantic embeddings, and abbreviation expansion to handle terminology variation across heterogeneous Earth Science repositories.

    \item We construct a literature-grounded benchmark from peer-reviewed publications and demonstrate that ReSearch consistently outperforms baseline methods, particularly for task-based queries expressing high-level research objectives.
\end{itemize}

\begin{figure}[htbp]
\centering
\includegraphics[width=0.6\textwidth]{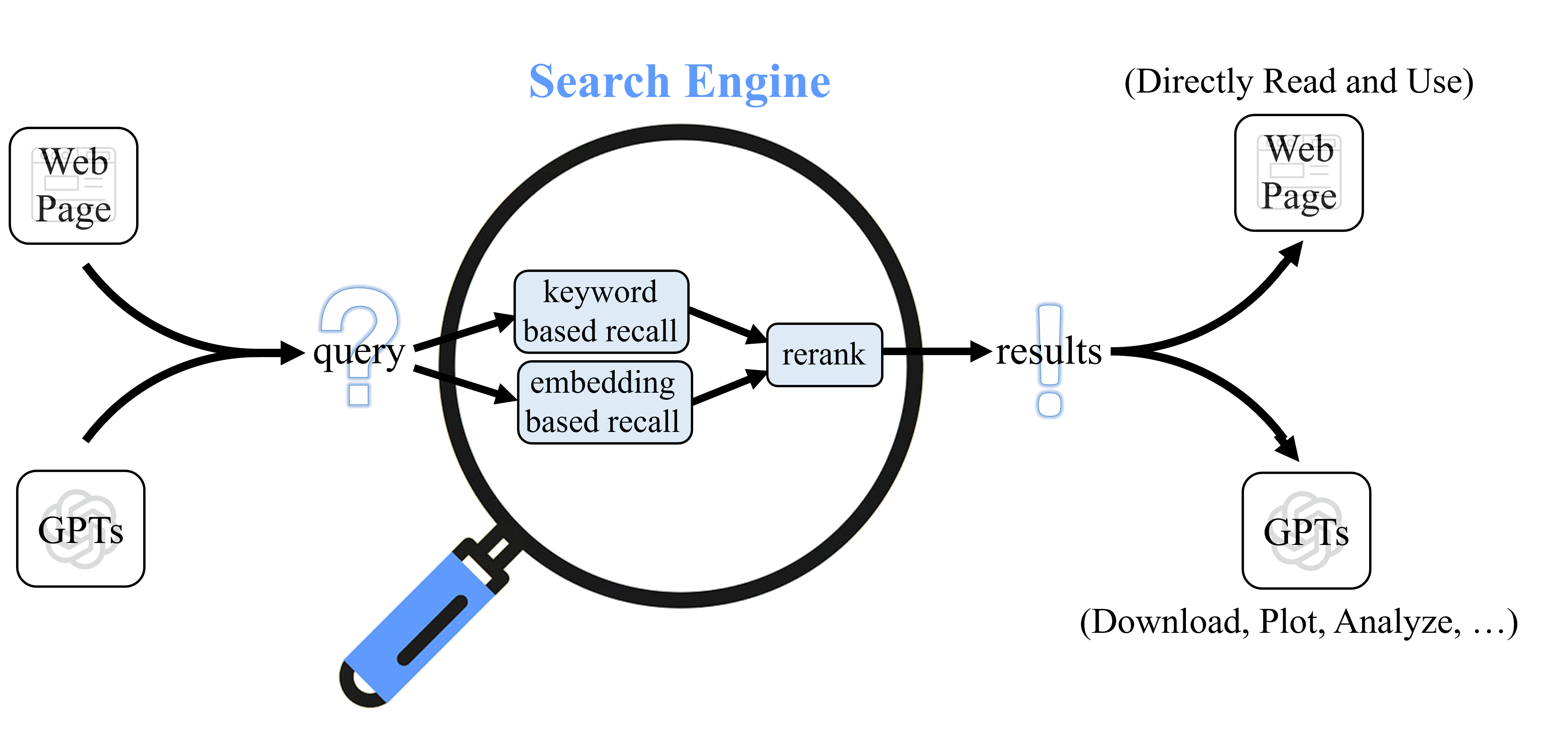}
\caption{Overview of the ReSearch framework. User queries enter the search engine from a web page or GPT-based agents. Inside the engine, keyword-based and embedding-based recall run in parallel, followed by LLM reranking. Results are returned for direct use or further processing by downstream agents.}
\label{fig:pipeline}
\end{figure}

\section{Related Work}

A growing body of work has explored the use of machine learning and semantic technologies to improve Earth Science data discovery and interoperability~\cite{addink2025, parisi2025}.
Major data portals and federated archives provide centralized access to observational and model datasets, while recent approaches have investigated semantic representations, embeddings, and knowledge graphs to bridge heterogeneity across repositories.
Although these systems improve data accessibility, they are primarily designed for human-driven retrieval and offer limited support for intent-aware or reasoning-oriented discovery.

AutoClimDS~\cite{autoclimds} integrates a knowledge graph with a multi-agent system to support climate data discovery.
It structures metadata as a bipartite graph linking datasets to auxiliary concepts (e.g., variables, keywords), enabling interpretable, one-hop traversal-based retrieval.
Dataset relevance is inferred indirectly via similarity to these auxiliary nodes.
While effective, this approach depends heavily on fixed taxonomies and lacks flexibility for iterative search or adaptation to diverse query intents.

Beyond knowledge-graph-based systems, embedding-based semantic retrieval has been employed to improve coverage across large, lexically diverse archives.
For example, Qiu et al.~\cite{qiu2019} proposed an ontology-enhanced embedding model to support keyphrase extraction from geoscience literature.
Ramachandran et al.~\cite{ramachandran2021} incorporated prediction-based embeddings to augment metadata in NASA Earth science repositories.
More recently, Weckmüller et al.~\cite{weckmuller2025} designed a multilingual embedding-based search framework for geo-textual datasets.
These approaches improve recall and language robustness, but often conflate retrieval and ranking, and offer limited mechanisms for integrating structured constraints or domain-specific reasoning.

Recent trends increasingly treat data discovery as a multi-stage reasoning process, where components such as intent interpretation, candidate generation, and contextual ranking are decoupled.
Such designs offer greater flexibility across heterogeneous repositories and can integrate with both semantic and agent-based systems.
This work builds on these directions, seeking to extend the reasoning capacity of discovery systems beyond rigid taxonomies and single-step retrieval.

\section{Problem Formulation}
\label{sec:problem}

The efficient discovery of relevant Earth Science data is impeded by a combination of structural and semantic barriers arising from decentralized archiving practices, heterogeneous metadata standards, and the use of specialized scientific terminology across subdisciplines.
These challenges hinder the translation of scientific questions into effective data queries and limit the scalability, reproducibility, and accessibility of Earth Science research workflows.

\begin{enumerate}
    \item \textbf{Metadata Heterogeneity.}
    Earth Science datasets are distributed across numerous repositories that employ inconsistent metadata schemas and naming conventions.
    Identical physical variables are often labeled differently across sources; for example, precipitation may appear as ``precip'', ``rainfall'', or ``pr'', while soil moisture, elevation, or land surface temperature exhibit similar inconsistencies.
    Such heterogeneity prevents direct matching through keyword-based search and necessitates additional semantic normalization before datasets can be meaningfully compared or integrated.

    \item \textbf{The Semantic Gap Between Scientific Intent and Metadata.}
    A persistent disconnect exists between high-level Earth Science research objectives (e.g., ``drought assessment'', ``flood risk analysis'', or ``cryospheric change detection'') and the low-level technical descriptors used in dataset metadata.
    Translating abstract scientific intent into specific variables, products, and spatiotemporal constraints requires substantial domain expertise.
    Even modern semantic search systems struggle to perform this translation reliably, often failing to infer the precise data requirements implied by a user's research goal.

    \item \textbf{Scale and Precision Trade-offs.}
    As Earth Science data archives grow to petabyte scales, achieving both high recall and high precision becomes increasingly difficult.
    Simple keyword matching frequently produces excessive noise, while overly restrictive filtering risks excluding relevant datasets.
    Effective discovery therefore requires nuanced reasoning over spatiotemporal coverage, variable semantics, and scientific context, capabilities that are largely absent from existing retrieval-centric systems.
\end{enumerate}

\paragraph{Problem Statement}
We formalize Earth Science data discovery as follows.
Let $q$ denote a natural language query expressing a scientific research objective, such as ``assessing the impact of Arctic sea ice loss on mid-latitude weather patterns.''
Let $\mathcal{D} = \{d_1, d_2, \ldots, d_n\}$ denote a heterogeneous corpus of dataset metadata records aggregated from multiple Earth Science repositories.
The goal is to produce a ranked list $\mathcal{R} = [(d_{i_1}, s_1), \ldots, (d_{i_k}, s_k)]$, where each $d_{i_j} \in \mathcal{D}$ is a candidate dataset and $s_j$ is a relevance score, such that datasets most pertinent to the research intent expressed in $q$ appear at the top.
The scores $s_j$ are used internally for ranking; in evaluation, we consider only the induced ordering of datasets.

The central difficulty lies in the semantic mismatch between $q$ and $\mathcal{D}$.
The query $q$ is typically expressed as a high-level scientific objective, while dataset metadata in $\mathcal{D}$ consists of low-level technical descriptors, variable names, and instrument identifiers.
Bridging this gap requires reasoning over domain knowledge, terminology variations, and implicit scientific context that is not explicitly encoded in either $q$ or $\mathcal{D}$.

\paragraph{Desiderata}
Given these challenges, an effective discovery system should satisfy three properties.
First, \emph{staged processing}: the translation from high-level intent to relevant datasets cannot be accomplished in a single retrieval step, so an effective system should decompose the discovery process into distinct stages, allowing each to address a specific aspect of the problem.
Second, \emph{hybrid retrieval}: no single retrieval method suffices for heterogeneous scientific metadata, as lexical matching captures exact terminology while semantic search handles vocabulary variation; an effective system should integrate multiple strategies to maximize coverage and minimize blind spots.
Third, \emph{recall-then-rank}: to balance coverage and precision, the system should first retrieve a broad candidate set, then apply context-aware ranking to promote the most relevant results, avoiding premature filtering that may exclude relevant datasets.

The ReSearch framework, introduced in the following section, is designed to satisfy these desiderata through a multi-stage architecture that combines lexical matching, semantic embedding retrieval, and large language model reasoning.

\section{Methodology}

\subsection{Data Sources}
To establish a solid basis for Earth Science research, we have integrated four primary data repositories.
These include the NASA Common Metadata Repository (CMR), NOAA OneStop, the Coupled Model Intercomparison Project Phase 6 (CMIP6), and the ERA5 (ECMWF Reanalysis v5) dataset.
These sources are a large and diverse collection of Earth observation and simulation data, ranging from satellite measurements to numerical model outputs.

Table \ref{tab:data_sources} summarizes the scale and characteristics of the integrated data sources.

\begin{table}[htbp]
\centering
\caption{Overview of Integrated Earth Science Data Sources}
\label{tab:data_sources}
\begin{tabular}{llrl}
\hline
\textbf{Data Source} & \textbf{Type} & \textbf{Datasets} & \textbf{Data Volume} \\
\hline
NASA CMR & Satellite Observations & $\approx$ 54,000 & PB scale \\
NOAA OneStop & Meteorological \& Oceanographic & $\approx$ 52,000 & Hundreds of TB \\
CMIP6 & Earth System Model Simulations & $\approx$ 102,000 & $\approx$ 20 PB \\
ERA5 & Reanalysis & 47 & $\approx$ 5 PB \\
\hline
\end{tabular}
\end{table}

A significant challenge in integrating these sources is the heterogeneity of metadata and variable naming conventions.
For example, precipitation is referenced by various identifiers such as ``precipitation\_rate'' or ``precipRate'' across different datasets.
To address this, our system employs metadata normalization and semantic mapping strategies to ensure consistent discovery capabilities.

\subsection{Search Strategy Design}
We propose a multi-stage search engine architecture tailored for Earth Science data discovery.
This architecture bridges the gap between natural language research queries and technical metadata schemas.
The pipeline consists of three stages: Query Understanding, Recall, and Reranking.

\paragraph{Stage 0. Query Understanding}
The initial stage focuses on intent classification and query refinement.
User queries are categorized into two types, Type A (specific data requests) and Type B (broad research goals).
While Type A queries undergo standard spell correction, Type B queries are processed using LLMs to translate abstract research objectives (e.g., ``flood analysis'') into specific, retrievable data requirements (e.g., ``precipitation'', ``storm surge'').
This stage also extracts structured constraints, such as temporal and spatial ranges, to aid downstream filtering.

\paragraph{Stage 1. Recall}
To maximize recall, we implement a hybrid retrieval strategy.
This involves a combination of structured filtering, utilizing the extracted spatiotemporal constraints, and a dual-path search mechanism.
We employ Best Matching 25 (BM25), a widely used term-frequency-based ranking function, for keyword-based matching to capture exact lexical matches in metadata fields.
Simultaneously, we utilize vector embedding search to identify semantically related datasets that may differ in terminology.
To further bridge the semantic gap, we apply abbreviation expansion, which augments the indexed metadata by inserting full-form expansions after detected abbreviations (e.g., ``MODIS'' $\rightarrow$ ``MODIS (Moderate Resolution Imaging Spectroradiometer)'').
This ensures a comprehensive candidate set that captures both explicit and implicit relevance.

\paragraph{Stage 2. Reranking}
The final stage refines the candidate set using an LLM-based reranker (GPT-4o).
The top $K_c$ ($K_c=50$) merged candidates from Stage 1 are formatted as a numbered list with truncated titles and summaries, then presented to the LLM with the user's query.
The LLM returns the indices of the $K_r$ ($K_r=10$) most relevant datasets in ranked order.
The final result is constructed by placing these $K_r$ datasets first, followed by the remaining candidates in their original order, so that recall at higher cutoffs is preserved while early precision is maximized.

\subsection{Evaluation Dataset Construction}
To rigorously evaluate the proposed search engine, we established a benchmark dataset derived from peer-reviewed Earth Science literature.
This approach ensures that our evaluation reflects authentic research needs and terminologies.

\begin{figure}[htbp]
\centering
\includegraphics[width=0.95\textwidth]{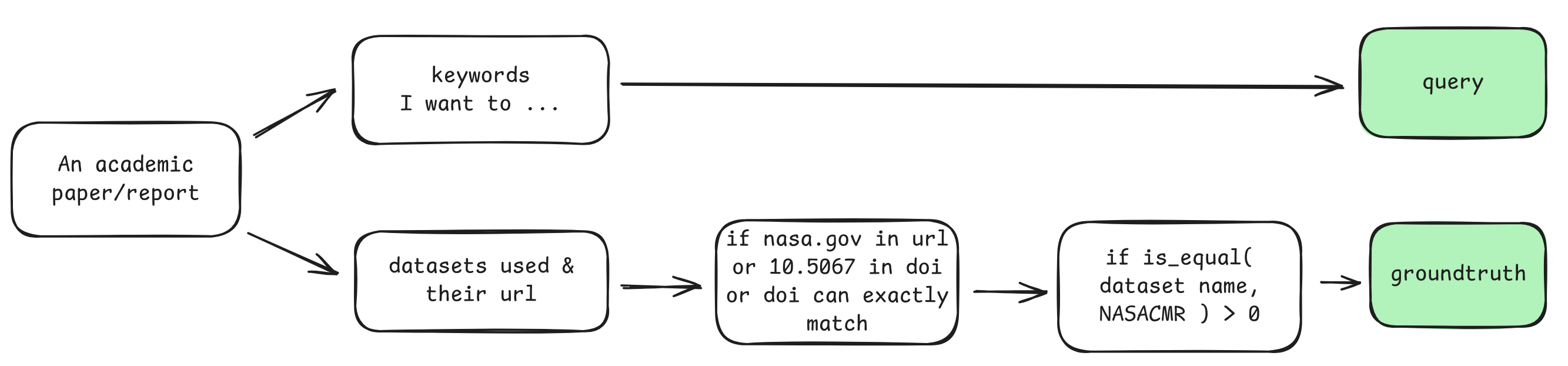}
\caption{Evaluation dataset construction pipeline. From academic papers, we extract queries and dataset references, then match datasets to NASA CMR entries to establish ground truth for retrieval evaluation.}
\label{fig:eval_pipeline}
\end{figure}

Figure \ref{fig:eval_pipeline} illustrates the construction pipeline for our evaluation dataset.
Starting from academic papers, we extract both research queries (keywords and ``I want to'' statements) and dataset references with their URLs.
Datasets are matched to NASA CMR entries through URL pattern matching and fuzzy name matching rules, producing query-groundtruth pairs for benchmark evaluation.

For each sampled paper, we utilized an LLM (GPT-4o) to extract the datasets explicitly cited by the authors. These datasets constitute the ground truth and were aligned with the integrated metadata repositories (e.g., NASA CMR) via unique short name identifiers.
To simulate diverse retrieval scenarios, we synthesized two distinct types of queries for each paper.
\begin{itemize}
    \item \textbf{Keyword-based Queries.} Aggregations of domain-specific keywords extracted from the text, simulating users searching with precise technical terminology.
    \item \textbf{Task-based Queries.} Natural language formulations (e.g., ``I want to analyze...'' statements) that describe high-level research objectives, representing users who may not be familiar with specific dataset identifiers.
\end{itemize}

Table \ref{tab:eval_stats} summarizes the evaluation dataset statistics.
Note that one paper was excluded from evaluation as none of its datasets matched entries in NASA CMR.
For keyword-based queries, keywords from each paper are aggregated into a single query (n=14 after exclusion).
For task-based queries, each paper contributes two ``I want to'' statements (n=28).

To evaluate generalization at larger scale, we constructed a second benchmark from the NASA Earth Observation Knowledge Graph~\cite{nasa_eo_kg_2024}, which contains 8,058 satellite dataset nodes and 138,704 publication nodes, of which 19,871 publications reference at least one dataset.
We ranked eligible publications by citation count, selected the top 1,000, and sampled every 10th paper to obtain 100 high-impact publications covering 206 dataset references.
Since the Knowledge Graph provides only titles and abstracts, keywords and ``I want to'' statements were generated by GPT-4o, consistent with the first benchmark.

\begin{table}[htbp]
\centering
\caption{Evaluation Dataset Statistics. NASA EO-KG refers to the NASA Earth Observation Knowledge Graph~\cite{nasa_eo_kg_2024}.}
\label{tab:eval_stats}
\begin{tabular}{lcc}
\hline
\textbf{Item} & \textbf{Literature} & \textbf{NASA EO-KG} \\
\hline
Sampled papers & 15 & 100 \\
Keyword-based queries & 14 & 100 \\
Task-based queries & 28 & 200 \\
Dataset references & 49 & 206 \\
\hline
\end{tabular}
\end{table}

\section{Experiments}

\subsection{Metrics}
We quantify system performance using three standard information retrieval metrics.
\begin{itemize}
    \item \textbf{Recall@K.} Measures the fraction of relevant datasets retrieved within the top $K$ results.
    For each query, Recall@K is computed as the number of ground truth datasets found within the top $K$ results divided by the total number of ground truth datasets for that query.
    The final score is the macro-average across all queries.
    \item \textbf{Mean Reciprocal Rank (MRR).} Evaluates the ranking effectiveness by calculating the average of the reciprocal ranks of the first relevant result. This metric reflects the system's ability to place a correct answer near the top of the list. It is defined as
    \[
    \mathrm{MRR}=\frac{1}{|Q|}\sum_{q \in Q}\frac{1}{\mathrm{rank}_q}
    \]
    where $Q$ is the set of queries and $\mathrm{rank}_q$ is the rank of the first relevant dataset for query $q$.
    \item \textbf{Mean Average Precision (MAP).} Assesses the overall quality of the ranking by considering the precision at each relevant dataset's position. For a query $q$, the average precision is defined as
    \[
    \mathrm{AP}(q) = \frac{1}{|\mathcal{G}_q|}
    \sum_{k=1}^{|\mathcal{R}_q|} P_q(k) \cdot rel_q(k),
    \]
    where $\mathcal{G}_q$ is the set of relevant datasets for query $q$,
    $\mathcal{R}_q$ is the ranked list returned by the system,
    $P_q(k)$ denotes the precision at rank $k$,
    and $rel_q(k) \in \{0,1\}$ indicates whether the item at rank $k$ is relevant.
    MAP is computed as the mean of AP over all queries.
\end{itemize}

\subsection{Main Results}

Table~\ref{tab:main_results} presents the retrieval performance of different methods on our literature-grounded benchmark.
We evaluate on two query types, keyword-based queries that use precise technical terminology, and task-based queries that express high-level research objectives.

\begin{table}[htbp]
\centering
\caption{Retrieval Performance on Earth Science Literature Benchmark}
\label{tab:main_results}
\begin{tabular}{lcccccc}
\hline
\textbf{Method} & \textbf{\RAt{10}} & \textbf{\RAt{20}} & \textbf{\RAt{50}} & \textbf{\RAt{100}} & \textbf{MRR} & \textbf{MAP} \\
\hline
\multicolumn{7}{l}{\textit{Keyword-based Queries (n=14)}} \\
AutoClimDS & 0.02 & 0.02 & 0.02 & 0.02 & 0.071 & 0.018 \\
BM25 & 0.12 & 0.15 & 0.27 & 0.32 & 0.040 & 0.033 \\
Embedding & 0.06 & 0.09 & 0.26 & 0.27 & 0.065 & 0.029 \\
Rerank & \textbf{0.18} & \textbf{0.28} & \textbf{0.33} & \textbf{0.38} & \textbf{0.172} & \textbf{0.121} \\
\hline
\multicolumn{7}{l}{\textit{Task-based Queries (n=28)}} \\
AutoClimDS & 0.01 & 0.01 & 0.01 & 0.01 & 0.071 & 0.013 \\
BM25 & 0.04 & 0.11 & \textbf{0.22} & 0.28 & 0.062 & 0.026 \\
Embedding & 0.07 & 0.12 & 0.19 & 0.23 & 0.049 & 0.031 \\
Rerank & \textbf{0.13} & \textbf{0.15} & 0.19 & \textbf{0.29} & \textbf{0.094} & \textbf{0.047} \\
\hline
\multicolumn{7}{l}{\textit{NASA EO-KG Benchmark -- Keyword-based Queries (n=100)}} \\
AutoClimDS & 0.00 & 0.00 & 0.00 & 0.00 & 0.000 & 0.000 \\
BM25 & 0.08 & 0.12 & 0.18 & 0.23 & 0.057 & 0.049 \\
Embedding & 0.09 & 0.14 & 0.18 & 0.24 & 0.057 & 0.047 \\
Rerank & \textbf{0.13} & \textbf{0.16} & \textbf{0.20} & 0.24 & \textbf{0.086} & \textbf{0.076} \\
\hline
\multicolumn{7}{l}{\textit{NASA EO-KG Benchmark -- Task-based Queries (n=200)}} \\
AutoClimDS & 0.00 & 0.00 & 0.00 & 0.00 & 0.000 & 0.000 \\
BM25 & 0.09 & 0.10 & 0.15 & 0.19 & 0.068 & 0.054 \\
Embedding & 0.09 & 0.14 & 0.18 & 0.20 & 0.061 & 0.051 \\
Rerank & \textbf{0.13} & \textbf{0.16} & 0.17 & \textbf{0.21} & \textbf{0.081} & \textbf{0.063} \\
\hline
\end{tabular}
\end{table}

\paragraph{AutoClimDS exhibits limited recall.}
AutoClimDS, which relies on knowledge graph traversal from auxiliary concept nodes, achieves extremely low recall across all cutoff thresholds ($\RAt{100} \leq 0.02$).
Although the relatively higher MRR (0.071) indicates that when AutoClimDS does retrieve a relevant dataset it tends to appear early in the ranked list, the overall coverage is severely insufficient for practical discovery scenarios.

\paragraph{ReSearch outperforms baselines, followed by BM25 and Embedding.}
For keyword-based queries, ReSearch achieves the best performance across all metrics ($\RAt{100} = 0.38$, $\mathrm{MRR} = 0.172$, $\mathrm{MAP} = 0.121$).
BM25 ranks second ($\RAt{100} = 0.32$), while Embedding ranks third ($\RAt{100} = 0.27$).
Notably, BM25 achieves higher recall than Embedding, indicating that exact lexical matching remains important for keyword queries; however, Embedding achieves higher MRR ($0.065$ vs.\ $0.040$), suggesting that semantic matching provides better ranking quality.
For task-based queries, the same ranking holds: ReSearch leads ($\RAt{100} = 0.29$, $\mathrm{MRR} = 0.094$), followed by BM25 ($\RAt{100} = 0.28$) and Embedding ($\RAt{100} = 0.23$).
The performance gap between methods is smaller than in keyword queries, but the ranking order remains consistent.

\paragraph{Task-based queries are more challenging.}
All methods exhibit lower performance on task-based queries compared to keyword-based queries.
ReSearch's $\RAt{100}$ drops from $0.38$ to $0.29$, BM25 from $0.32$ to $0.28$, and Embedding from $0.27$ to $0.23$.
Task-based queries express high-level research objectives (e.g., ``I want to study the impact of drought on agriculture''), which creates a semantic gap with dataset metadata.
Bridging this gap is the core challenge that our multi-stage framework addresses.

\subsection{Hybrid retrieval improves recall and ranking}

To quantify the benefit of combining BM25 and embedding retrieval, we compare retrieval performance across three configurations: BM25 only, Embedding only, and BM25+Embedding fusion.
Table~\ref{tab:ablation_fusion} presents the results.

\begin{table}[htbp]
\centering
\caption{Effect of Retriever Fusion (without reranking). Combining BM25 and Embedding retrieval improves recall at higher cutoffs.}
\label{tab:ablation_fusion}
\begin{tabular}{lcccccc}
\hline
\textbf{Method} & \textbf{\RAt{10}} & \textbf{\RAt{20}} & \textbf{\RAt{50}} & \textbf{\RAt{100}} & \textbf{MRR} & \textbf{MAP} \\
\hline
\multicolumn{7}{l}{\textit{Keyword-based Queries (n=14)}} \\
BM25 & \textbf{0.12} & 0.15 & 0.27 & 0.32 & 0.040 & 0.033 \\
Embedding & 0.06 & 0.09 & 0.26 & 0.27 & 0.065 & 0.029 \\
BM25+Embedding & 0.05 & \textbf{0.18} & \textbf{0.33} & \textbf{0.38} & \textbf{0.098} & \textbf{0.043} \\
\hline
\multicolumn{7}{l}{\textit{Task-based Queries (n=28)}} \\
BM25 & 0.04 & 0.11 & \textbf{0.22} & 0.28 & 0.062 & 0.026 \\
Embedding & \textbf{0.07} & 0.12 & 0.19 & 0.23 & 0.049 & 0.031 \\
BM25+Embedding & 0.06 & \textbf{0.15} & 0.19 & \textbf{0.29} & \textbf{0.094} & \textbf{0.047} \\
\hline
\end{tabular}
\end{table}

Combining BM25 and embedding retrieval yields substantial improvements in both recall and ranking quality.
For keyword-based queries, hybrid retrieval achieves $\RAt{100} = 0.38$ compared to $0.32$ for BM25 alone and $0.27$ for embedding alone.
MRR improves from $0.040$ (BM25) and $0.065$ (Embedding) to $0.098$, indicating that relevant datasets are ranked higher when both retrieval signals are combined.
Similar patterns hold for task-based queries, where hybrid retrieval achieves the highest MRR ($0.094$) and the best $\RAt{100}$ ($0.29$ vs.\ $0.28$ for BM25 and $0.23$ for Embedding).
The improvement demonstrates that BM25 and embedding-based methods capture complementary aspects of relevance: BM25 excels at exact lexical matching, while embeddings capture semantic similarity.
By prioritizing datasets retrieved by both methods and interleaving the remaining results, the hybrid approach benefits from both signals.

\subsection{Ablation Study: Effect of LLM Reranking}

To quantify the benefit of LLM-based reranking, we fix the retriever to BM25+Embedding and compare performance with and without reranking.
Table~\ref{tab:ablation_rerank} presents the results.

\begin{table}[htbp]
\centering
\caption{Effect of LLM Reranking (with BM25+Embedding retrieval). Reranking improves early precision without affecting recall.}
\label{tab:ablation_rerank}
\begin{tabular}{lcccccc}
\hline
\textbf{Method} & \textbf{\RAt{10}} & \textbf{\RAt{20}} & \textbf{\RAt{50}} & \textbf{\RAt{100}} & \textbf{MRR} & \textbf{MAP} \\
\hline
\multicolumn{7}{l}{\textit{Keyword-based Queries (n=14)}} \\
Without Reranking & 0.05 & 0.18 & 0.33 & 0.38 & 0.098 & 0.043 \\
With Reranking & \textbf{0.18} & \textbf{0.28} & 0.33 & 0.38 & \textbf{0.172} & \textbf{0.121} \\
\hline
\multicolumn{7}{l}{\textit{Task-based Queries (n=28)}} \\
Without Reranking & 0.06 & 0.15 & 0.19 & 0.29 & 0.094 & 0.047 \\
With Reranking & \textbf{0.13} & 0.15 & 0.19 & 0.29 & 0.094 & 0.047 \\
\hline
\end{tabular}
\end{table}

For keyword-based queries, reranking yields dramatic improvements: $\RAt{10}$ increases from $0.05$ to $0.18$, $\RAt{20}$ from $0.18$ to $0.28$, MRR from $0.098$ to $0.172$, and MAP from $0.043$ to $0.121$.
This demonstrates that the LLM can effectively identify the most relevant datasets from the candidate pool and promote them to the top of the ranked list.
Since users typically examine only the top 10--20 results, this improvement in early precision is critical for user experience.
Meanwhile, $\RAt{50}$ and $\RAt{100}$ remain unchanged ($0.33$ and $0.38$, respectively), which is expected since reranking only reorders the top-50 candidates without introducing new datasets.

For task-based queries, $\RAt{10}$ improves from $0.06$ to $0.13$, but $\RAt{20}$, MRR, and MAP remain unchanged.
This suggests that for queries expressing high-level research objectives, even LLM reasoning cannot fully bridge the semantic gap between query intent and dataset metadata.
Nevertheless, the improvement in $\RAt{10}$ indicates that reranking still provides some benefit in surfacing the most relevant results at the very top of the list.

\subsection{Ablation Study: Effect of Abbreviation Expansion}

To quantify the contribution of abbreviation expansion, we compare retrieval performance with and without this preprocessing step.
Abbreviation expansion augments the indexed metadata by inserting full-form expansions after detected abbreviations (e.g., ``MODIS'' $\rightarrow$ ``MODIS (Moderate Resolution Imaging Spectroradiometer)''), enabling retrieval systems to match queries that use either the abbreviation or the full term.

Table~\ref{tab:ablation_abbr} presents the results on keyword-based queries, where the effect is most pronounced.

\begin{table}[htbp]
\centering
\caption{Effect of Abbreviation Expansion on Keyword-based Queries (n=14)}
\label{tab:ablation_abbr}
\begin{tabular}{lcccccc}
\hline
\textbf{Method} & \textbf{\RAt{10}} & \textbf{\RAt{20}} & \textbf{\RAt{50}} & \textbf{\RAt{100}} & \textbf{MRR} & \textbf{MAP} \\
\hline
BM25 & 0.05 & 0.05 & 0.18 & 0.31 & 0.025 & 0.020 \\
BM25 + AbbrExp & \textbf{0.12} & \textbf{0.15} & \textbf{0.27} & \textbf{0.32} & \textbf{0.040} & \textbf{0.033} \\
\hline
Embedding & 0.02 & 0.02 & 0.22 & 0.25 & 0.052 & 0.022 \\
Embedding + AbbrExp & \textbf{0.06} & \textbf{0.09} & \textbf{0.26} & \textbf{0.27} & \textbf{0.065} & \textbf{0.029} \\
\hline
\end{tabular}
\end{table}

\paragraph{Abbreviation expansion substantially improves early recall.}
For BM25, abbreviation expansion increases $\RAt{10}$ from 0.05 to 0.12 (140\% relative improvement) and $\RAt{20}$ from 0.05 to 0.15 (200\% improvement).
Similarly, embedding retrieval benefits from expansion, with $\RAt{10}$ improving from 0.02 to 0.06 (200\% improvement).
These gains demonstrate that abbreviation mismatch is a significant source of retrieval failure in Earth Science data discovery, where datasets are frequently described using instrument acronyms, mission abbreviations, and technical shorthand.

\paragraph{The effect diminishes at higher cutoffs.}
At $\RAt{100}$, the improvement from abbreviation expansion is more modest (BM25: $0.31 \rightarrow 0.32$; Embedding: $0.25 \rightarrow 0.27$).
This pattern suggests that abbreviation expansion primarily helps surface relevant datasets earlier in the ranked list, rather than recovering entirely new relevant datasets.
The practical implication is that abbreviation expansion is particularly useful for user-facing applications where users examine only the top results.

\section{Conclusion}

In this paper, we presented \textbf{ReSearch}, a multi-stage framework for Earth Science data discovery that addresses a foundational bottleneck in data-driven research, namely translating high-level scientific intent into reliable and scalable access to heterogeneous data repositories.
By reformulating data discovery as an iterative reasoning process, ReSearch explicitly separates intent interpretation, high-recall retrieval, and context-aware ranking, enabling effective search across large and inconsistently described Earth Science archives.

The proposed framework integrates complementary techniques from information retrieval and machine learning, including lexical matching, semantic embedding search, and abbreviation expansion, with a modular architecture designed to support large language model reasoning in downstream ranking stages.
This hybrid design improves tolerance of metadata heterogeneity and linguistic variation, which are pervasive challenges in Earth Science data ecosystems.
Through a literature-grounded evaluation derived from peer-reviewed studies, we demonstrated that the retrieval components of ReSearch outperform baseline methods, particularly for task-based queries that reflect abstract scientific objectives rather than dataset-specific terminology.

Beyond empirical performance gains, this work highlights the importance of treating data discovery as a first-class component of scientific workflows.
Errors or omissions at the discovery stage can propagate downstream, affecting analysis quality, reproducibility, and scientific conclusions.
By improving the reliability and transparency of discovery, ReSearch supports more reproducible and inclusive Earth Science research, lowering barriers for interdisciplinary and data-driven investigations.

While this study focuses on Earth Science data repositories, the ReSearch framework is domain-agnostic and applicable to other scientific fields characterized by large-scale, heterogeneous datasets and evolving terminology.
Future work will explore extensions of ReSearch to support the discovery of analysis methods, modeling workflows, and simulation strategies, as well as tighter integration with knowledge-graph-based and agentic systems for end-to-end automated research pipelines.
These directions suggest potential for ReSearch to serve as a reusable search component within broader scientific discovery workflows.

\section*{Acknowledgments}
HY was partially supported by the US National Science Foundation under awards IIS-2520978, GEO/RISE-5239902, the Office of Naval Research Award N00014-23-1-2007, DOE (ASCR) Award DE-SC0026052, and the DARPA D24AP00325-00.
Approved for public release; distribution is unlimited.

%\section*{Conflict of Interest Statement}
%The authors have no conflicts of interest to disclose.

% \bibliographystyle{plain}
\bibliographystyle{apalike}
\bibliography{main}

\appendix
\section{Evaluation Dataset Details}
\label{appendix:eval_data}

Table~\ref{tab:eval_datasets} presents the datasets extracted from the sampled Earth Science papers that were successfully matched to entries in NASA CMR.
For each paper, we list the dataset names as they appear in the original publication, along with the number of matching entries found in our integrated metadata repository.

The \textbf{Matches} column indicates how many NASA CMR dataset entries correspond to each extracted dataset name.
This count is determined by soft matching, which accounts for variations in naming conventions, version numbers, and formatting differences between how datasets are cited in papers and how they are catalogued in NASA CMR.
A high match count (e.g., 1113 for ``MODIS'') indicates that the paper referenced a broad instrument or mission name rather than a specific data product, resulting in many candidate datasets.
Conversely, a low match count (e.g., 1 for ``GPM IMERG Final Precipitation L3 1 day 0.1 degree x 0.1 degree V06'') indicates a precise product reference that resolves to a unique or near-unique entry.

This variation in specificity reflects the diversity of citation practices in Earth Science literature.
Some papers cite datasets at the mission or instrument level, while others provide exact product identifiers.
Our evaluation treats all matched entries as valid ground truth, acknowledging that a retrieval system should be able to surface any of the matching products when a user queries with the dataset name as it appears in the source paper.

{\small
\begin{longtable}{p{9.5cm}r}
\caption{Datasets Extracted from Evaluation Papers (Matched Only)} \label{tab:eval_datasets} \\
\hline
\textbf{Dataset} & \textbf{Matches} \\
\hline
\endfirsthead
\multicolumn{2}{c}{\tablename\ \thetable\ -- continued from previous page} \\
\hline
\textbf{Dataset} & \textbf{Matches} \\
\hline
\endhead
\hline \multicolumn{2}{r}{Continued on next page} \\
\endfoot
\hline
\endlastfoot
% Paper 1
\multicolumn{2}{l}{\textit{Global high-resolution drought indices for 1981--2022}} \\
\quad Normalized Difference Vegetation Index (NDVI) from MODIS (MOD13C2) & 86 \\
\quad Land surface temperature data from MODIS (MOD11C3) & 3 \\
\hline
% Paper 2
\multicolumn{2}{l}{\textit{Global scale assessment of urban precipitation anomalies}} \\
\quad GPM IMERG Final Precipitation L3 1 day 0.1 degree x 0.1 degree V06 & 1 \\
\quad Level-3 Aura/OMI Global Aerosol Data (OMAEROe) & 2 \\
\quad MOD11C3 MODIS/Terra Land Surface Temperature Monthly L3 Global 0.05Deg CMG V006 & 3 \\
\quad MCD12C1 MODIS/Terra+Aqua Land Cover Type Yearly L3 Global 0.05Deg CMG & 2 \\
\quad Global Human Settlement Layer: Population and Built-Up Estimates & 1 \\
\hline
% Paper 3
\multicolumn{2}{l}{\textit{Regional-scale intelligent optimization and topography impact in restoring global precipitation data gaps}} \\
\quad Integrated Multi-satellite Retrievals for GPM-Early (IMERG-Early) & 2 \\
\hline
% Paper 4
\multicolumn{2}{l}{\textit{Regional analysis of the 2015--16 Lower Mekong River basin drought}} \\
\quad GPM IMERG Final Precipitation L3 1 month 0.1 degree x 0.1 degree V06 & 1 \\
\quad GRACE RL06 & 38 \\
\hline
% Paper 5
\multicolumn{2}{l}{\textit{Tropical deforestation causes large reductions in observed precipitation}} \\
\quad GPCP & 14 \\
\quad GPM & 476 \\
\quad MERRA-2 & 109 \\
\quad TRMM & 120 \\
\hline
% Paper 6
\multicolumn{2}{l}{\textit{The application of satellite sensors in hydrology for water resource management}} \\
\quad TMPA & 12 \\
\quad GPM/IMERG & 9 \\
\quad MODIS & 1113 \\
\quad VIIRS & 854 \\
\quad SMAP & 203 \\
\quad ASTER & 551 \\
\quad ECOSTRESS & 47 \\
\quad GRACE & 80 \\
\quad Topex/Poseidon & 6 \\
\hline
% Paper 7
\multicolumn{2}{l}{\textit{Evaluation of a MODIS triangle-based algorithm for evapotranspiration estimates}} \\
\quad MODIS MYD11A2 & 2 \\
\quad MODIS MYD07 & 3 \\
\quad MODIS MYD06 & 3 \\
\quad MODIS MOD13Q1 & 2 \\
\quad MODIS MYD13Q1 & 2 \\
\quad MODIS MYD03 & 3 \\
\quad MODIS MYD05\_L2 & 3 \\
\quad MODIS MYD06\_L2 & 3 \\
\quad MODIS MYD07\_L2 & 3 \\
\quad MODIS MYD11\_L2 & 4 \\
\quad MODIS MYD11A1 & 2 \\
\hline
% Paper 8
\multicolumn{2}{l}{\textit{Satellite-based water use dynamics using historical Landsat data (1984--2014)}} \\
\quad Landsat 5 & 30 \\
\quad Landsat 7 & 18 \\
\quad Landsat 8 & 4 \\
\hline
% Paper 9
\multicolumn{2}{l}{\textit{Satellite-based remote sensing data set of global surface water storage change}} \\
\quad Global Lake/Reservoir Surface Inland Water Height GREALM V.2 & 1 \\
\quad Global Lake/Reservoir Surface Inland Water Area Extent V2 & 1 \\
\quad Lake and Reservoir Storage Time Series V2 & 1 \\
\hline
% Paper 10
\multicolumn{2}{l}{\textit{Satellite observed recent rising water levels of global lakes and reservoirs}} \\
\quad ICESat GLAH14 & 1 \\
\hline
% Paper 11
\multicolumn{2}{l}{\textit{Satellites reveal widespread decline in global lake water storage}} \\
\quad Modern-Era Retrospective analysis for Research and Applications version 2 (MERRA-2) & 109 \\
\hline
% Paper 13
\multicolumn{2}{l}{\textit{GloLakes: water storage dynamics for 27000 lakes globally from 1984 to present}} \\
\quad ATLAS/ICESat-2 L3A Along-Track Inland Surface Water Data, Version 6 (ATL13) & 3 \\
\quad ATLAS/ICESat-2 L3A Along Track Inland Surface Water Data Quick Look, Version 6 (ATL13QL) & 1 \\
\quad Global Reservoirs and Lakes Monitor (GREALM) & 1 \\
\hline
% Paper 14
\multicolumn{2}{l}{\textit{HydroSat: geometric quantities of the global water cycle from geodetic satellites}} \\
\quad MODIS MOD09Q1 & 2 \\
\hline
% Paper 15
\multicolumn{2}{l}{\textit{An evaluation of ERA5 precipitation for climate monitoring}} \\
\quad Tropical Rainfall Measuring Mission TRMM/3B43 & 1 \\
\hline
\end{longtable}
}

\section{Prompt Templates}
\label{appendix:prompts}

This section presents the prompt templates used in the ReSearch framework for LLM-based components.

\subsection{Intent Classification Prompt}

This prompt classifies user queries into Type A (specific data requests) or Type B (broad research goals).

\begin{verbatim}
You are an Earth science data search assistant. Classify the user's query intent.

**Type A**: The user knows what data they want. They mention specific:
- Data variables (e.g., precipitation, temperature, sea level)
- Platforms/satellites (e.g., GPM, MODIS, Landsat)
- Dataset names (e.g., ERA5, CMIP6)
- Data formats or sources

**Type B**: The user describes a research goal but doesn't know what data
they need. They mention:
- Research topics (e.g., "study flooding", "ocean acidification effects")
- Problems to solve (e.g., "predict drought", "analyze hurricanes")
- General questions without specific data mentions

Examples:
- "percipitation data GPM" -> Type A (mentions specific variable and platform)
- "I want to study Florida flooding" -> Type B (describes research goal)
- "ERA5 temperature 2020" -> Type A (mentions dataset name)
- "how to predict drought in Africa" -> Type B (describes problem to solve)

User query: {query}

Respond with ONLY "A" or "B", nothing else.
\end{verbatim}

\subsection{Query Rewrite Prompt}

This prompt converts high-level research goals into data-oriented search queries for Type B queries.

\begin{verbatim}
You are an Earth science data search expert. The user has a research goal
but doesn't know what specific data they need.

Your task: Convert the research goal into a data-oriented search query
that will help find relevant Earth science datasets.

Think about:
1. What Earth science variables are relevant to this research?
   (e.g., precipitation, temperature, wind, humidity)
2. What phenomena or measurements are needed?
   (e.g., extreme events, trends, anomalies)
3. What spatial/temporal aspects matter?
   (e.g., global, regional, daily, monthly)

User's research goal: {query}

Return a JSON object with two fields:
- "reasoning": Your brief analysis of what data the user needs (2-3 sentences)
- "query": A concise, data-oriented search query in English

Example response:
{"reasoning": "The user wants to study flooding in Florida. This requires
precipitation data (especially extreme rainfall), storm surge data, and
sea level data for the Florida/Gulf of Mexico region.",
"query": "precipitation extreme rainfall storm surge sea level Florida
Gulf of Mexico"}

Return ONLY the JSON object, no other text.
\end{verbatim}

\subsection{Reranking Prompt}

This prompt instructs the LLM to rerank candidate datasets by relevance to the user's query.

\begin{verbatim}
You are an expert Earth science dataset retrieval reranker.

Given a user query and numbered candidate datasets, rank them by relevance.

Rules:
- Prefer datasets that directly measure the key variables/phenomena
  implied by the query.
- Only return numbers from the given candidates. Do NOT invent new numbers.
- Return the TOP {top_n} most relevant candidates only.

Return JSON: {"ranked": [3, 1, 5, ...]}

Query: {query}

Candidates:
{candidates_text}
\end{verbatim}

\subsection{Paper Information Extraction Prompt}

This prompt extracts dataset references, keywords, and research objectives from academic papers to construct the evaluation benchmark.

\begin{verbatim}
You are a research paper analyst. Your task is to extract specific
information from academic papers.

Respond ONLY with a valid JSON object (no markdown, no explanation)
with these exact keys:
{
  "datasets": [{"name": "...", "doi_or_url": "..."}, ...],
  "keywords": [...],
  "is_original_keywords": true/false,
  "i_want_to": [...]
}

Instructions:
1. **datasets**: List ALL datasets used in the paper. Be EXACT with names.
   - Include version numbers, temporal ranges, spatial resolution if mentioned
   - Examples: "GPM IMERG Final Precipitation L3 1 month 0.1 degree x
     0.1 degree V06", "GPM_3IMERGM"
   - Do NOT include generic terms like "satellite data"
   - Only include actual downloadable/accessible data products

2. **keywords**: List the paper's keywords.
   - If the paper has an explicit "Keywords:" section, use those EXACTLY
   - If no explicit keywords exist, extract the 5 most important
     topical keywords

3. **is_original_keywords**: Boolean.
   - true if keywords came from the paper's explicit "Keywords:" section
   - false if you extracted them yourself

4. **i_want_to**: List 2 statements in "I want to ..." format.
   - Write from the author's perspective BEFORE writing the paper
   - Describe what they wanted to achieve/investigate/develop
   - Example: "I want to develop a machine learning model to predict
     urban heat island intensity using satellite observations"
\end{verbatim}

\end{document}